\documentclass[aas2pp4]{aastex}
\usepackage{isolatin1}
\usepackage{natbib}

\slugcomment{To appear in Astrophysical Journal Letters}

\shorttitle{{\sc [OII]}$\lambda$3727 local luminosity function}
\shortauthors{Gallego et al.}

\begin{document}


\title{The {\sc [OII]}$\lambda$3727 Luminosity Function of the Local
Universe\altaffilmark{1}}


\author{J. Gallego, C.E. Garc\'{\i}a-Dab\'o, 
        J. Zamorano, A. Arag\'{o}n--Salamanca\altaffilmark{2}, M. Rego} 
\email{jgm@astrax.fis.ucm.es}
\affil{Departamento de Astrof\'{\i}sica, Universidad
       Complutense de Madrid}
\affil{Facultad CC F\'{\i}sicas, Ciudad Universitaria, E-28040 Madrid, Spain}


\altaffiltext{1}{Partly based on observations collected at the
German-Spanish Astronomical Center, Calar Alto, Spain, operated by the
Max-Planck-Institute f\"{u}r Astronomie (MPIA), Heidelberg, jointly
with the Spanish National Commission for Astronomy. Partly based on
observations made with the Isaac Newton Telescope operated on the
island of La Palma by the Royal Greenwich Observatory in the Spanish
Observatorio del Roque de Los Muchachos of the Instituto de
Astrof\'{\i}sica de Canarias.}  
\altaffiltext{2}{School of Physics \& Astronomy,
University of Nottingham, Nottingham NG7 2RD, England.}


\begin{abstract}
The measurement of the Star Formation Rate density of the Universe is
of prime importance in understanding the formation and evolution of
galaxies.  The {\sc [OII]}$\lambda$3727 emission line flux, easy to
measure up to z$\approx$1.4 within deep redshift surveys in the
optical and up to z$\approx$5.4 in the near infrared, offers a
reliable means of characterizing the star formation properties of
high-$z$ objects. In order to provide the high-z studies with a local
reference, we have measured total {\sc [OII]}$\lambda$3727 fluxes for
the well analyzed local sample of star-forming galaxies from the
Universidad Complutense de Madrid Survey. This data is used to derive
the {\sc [OII]}$\lambda$3727 luminosity function for local
star-forming galaxies. When compared with similar luminosity densities
published for redshift up to z$\approx$1, the overall evolution
already observed in the star formation activity of the Universe is
confirmed.
\end{abstract}


\section{Introduction}
The measurement of the Star Formation Rate (SFR) density of the
Universe as a function of look-back time is a fundamental parameter in
order to understand the formation and evolution of galaxies.  The
current picture, outlined in the last years, is that the global SFR
density has been declining from a peak at redshift of z$\sim$1.5 to
the present day value \citep[see][and references
therein]{astro-ph/0105280}.  Despite one of the best direct
measurements in the optical of current SFR is the nebular H$\alpha$
6563\AA\ luminosity \citep{1998ARA&A..36..189K,2001MNRAS.323..887C},
this emission line is only detectable with CCDs out to z$\approx$0.4.
H$\alpha$ luminosities have been used to trace the SFR density in this
redshift range
\citep{1995ApJ...455L...1G,1998ApJ...495..691T,2001ApJ...550..593J,2001A&A...379..798P}.

The alternative indicators used in the optical for estimating the SFR
 at high$-z$ have been mainly two: UV continuum luminosities
 \citep{1996ApJ...460L...1L,1996MNRAS.283.1388M,1997sfnf.conf..481M,1997ApJ...486L..11C};
 and {\sc [OII]}$\lambda$3727 luminosities
 \citep{1997ApJ...481...49H,1998ApJ...504..622H}.
 UV fluxes are easy to obtain from broad-band photometry and, when
 combined with photometric redshifts, are useful to analyze large
 samples.  The {\sc [OII]} flux is easy to measure when in emission
 and is available in the optical until $z\simeq1.5$.

In this paper we use {\sc [OII]}$\lambda$3727 fluxes for the
Universidad Complutense de Madrid local sample to derive
the local {\sc [OII]}$\lambda$3727 luminosity function.  In \S 2 we
discuss the sample and the new data. In \S 3, the local {\sc [OII]}
luminosity function is obtained.  Finally, in \S 4 we compare the
local luminosity density with estimations already available for higher
redshifts.  A Friedman cosmology with H$_0 = 100$ km s$^{-1}$
Mpc$^{-1}$ and q$_0 = 0.5$ has been used.

\section{The UCM survey of Local Star-forming Galaxies: {\sc [OII]}$\lambda$3727 data} 
The Universidad Complutense de Madrid (hereafter UCM) objective-prism
survey \citep[List I\&II,][]{1994ApJS...95..387Z,1996ApJS..105..343Z}
provides an ideal tool for studying the population and properties of
star-forming galaxies at low redshift.  The sample consists of 191
galaxies in 471.4 deg$^2$ with z$\lesssim$0.045 and equivalent width
EW(H$\alpha$+[NII])$\gtrsim$20\AA.  Objects were selected according to
their H$\alpha$+[NII] + continuum fluxes and therefore they constitute a
representative sample of current star-forming galaxies.  Accurate
spectrophotometry \citep{1996A&AS..120..323G,1997ApJ...475..502G} and
broad-band photometry in the optical
\citep{1996A&AS..118....7V,1996A&AS..120..385V,2000A&AS..141..409P,2001A&A...365..370P} have already been published. 

In September 1996 a new spectroscopic run was carried out to obtain
high quality data in the {\sc [OII]}$\lambda$3727 region.  In this run
a total of 108 (56\%) UCM galaxies were observed again. The telescope
used was the 2.5-m Isaac Newton Telescope (INT) at La Palma (Spain).
The wavelength range covered was 3640\AA-6180\AA \ with a dispersion
of 2.5\AA/pixel.  The slit width was 2 arcsec (spectral resolution of
$\sim$7\AA) except for 13 galaxies that, given their large size, were observed
with a 4 arcsec slit.  The reduction was made following the standard
procedure using both IRAF and the Reduceme software package
\citep{1999PhDT........12C}.

When joining the new results to the \citet{1996A&AS..120..323G} data,
a final 92\% (176 out of 191) of the UCM objects had the {\sc
[OII]}$\lambda$3727 region well covered.  In 145 of these 191 objects
(76\%), we have measured {\sc [OII]} in emission with EW$\gtrsim$5\AA.
A total fraction of 24\% of objects show no line. Most of them are
starburst nuclei-like objects, with a contribution to the global SFR
(as measured from H$\alpha$) that is $\sim$25\% of the total amount.
The quality of the new spectra revealed the existence of underlying
absorptions in the H$\beta$ emission line.  However, we could not
estimate Balmer decrements because the H$\alpha$ region was not
covered. Color excesses were re-calculated using the observed
H$\alpha$/H$\beta$ from \citet{1996A&AS..120..323G} but assuming
equivalent widths of the Balmer lines in absorption equal to 3\AA\
\citep{1999ApJS..125..489G}.  The new values, 0.2 in average smaller
than previous ones, were used to correct for reddening. The H$\gamma$ line,
when observed, provided a consistency check.  We were
not able to estimate extinction for five galaxies.  

Based on the emission line diagnostic diagrams and morphological
properties, \citet{1996A&AS..120..323G} classified spectroscopically
each of the UCM galaxies. We decided to remove all UCM galaxies under
the Seyfert type. Independently of its spectroscopic class, each
nucleated galaxy could harbor a dwarf seyfert
nucleus. \citet{1997ApJ...487..568H} estimated that 43\% of galaxies
had such a nucleus with H$\alpha$ luminosity 1/100th of the typical
H$\alpha$ luminosity of a Seyfert galaxy.  This amount turns out to be
about one hundredth of the L$^*$ measured by
\citet{1995ApJ...455L...1G} We decided not to include any correction
relative to this effect. Once the AGNs and objects with no emission
were removed, the final sample consisted of 134 and 129 galaxies with
observed and extinction corrected luminosity.
\section{The Local {\sc [OII]}$\lambda$3727 luminosity function}
Direct information on the amount and nature of the present-day SFR in
the local Universe can be obtained by constructing the {\sc
[OII]}$\lambda$3727 luminosity function for galaxies with current star
formation activity.  To correct for the signal not covered by the
slit, the total {\sc [OII]}$\lambda$3727 luminosity line for each
galaxy was computed from the {\sc [OII]} equivalent width and its
continuum as obtained from the Johnson B band magnitude. Given that
the {\sc [OII]}$\lambda$3727 line is not included into the B band for
the UCM galaxies, we analyzed the ratio of the actual continuum
adjacent to the {\sc [OII]}$\lambda$3727 line, and the average
continuum within the B band, as measured in our spectra. The average
$k=F^c_B / F^c_{[OII]}$ ratios for each spectroscopic type (0.94, 1.15
and 1.43 for Blue Compact Dwarfs, HII-like and disk-like objects
respectively) were applied.
The observed total [OII] line luminosity $L_{obs}^{OII}$ was
computed as
\begin{eqnarray}
 L_{obs}^{OII}    & = & EW_{obs}^{OII} \; L_{obs}^{c,OII} 
\end{eqnarray}
and the adjacent continuum $L_{obs}^{c,OII}$ as the scaled B continuum 
$L_{obs}^{c,B}$ without the line contamination
\begin{eqnarray}
 L_{obs}^{c,OII}  & = &  L_{obs}^{c,B} \; / \; (k + EW_{obs}^{OII} \: 
\frac{Qe(3727)}{\Delta \lambda(B)}) \nonumber \\
 L_{obs}^{c,B}   & = & P_{0}^{B} \; 10^{-0.4 \: m_B} \; D(z) 
\end{eqnarray}
where $m_B$ is the Johnson B apparent magnitude, $P_{0}^{B}$ the
corresponding photometric zero point and $D(z)$ is the luminosity
distance. It is worth noting that the B band magnitudes were not corrected
for galactic extinction in order to compare with those observed [OII] 
luminosities measured from not extinction-corrected spectroscopy.
The mean galactic extinction in the B band for the UCM sample is
0.23$\pm$0.21 if \citet{1998ApJ...500..525S} is considered
or 0.12$\pm$0.10 if \citet{1982AJ.....87.1165B} is considered.

In a sample such as the UCM survey, the completeness is determined by
the H$\alpha$+[NII] line+continuum flux. This is a pseudo apparent
magnitude proportional to the current star-forming activity of the
source.  \citet{1995ApJ...455L...1G} used the V/Vmax test to obtain a
complete sample of 176 galaxies within the UCM survey. These objects
were selected with line+continuum flux larger than $1.9 \times
10^{-14}\,$erg$\,$cm$^{-2}\,$s$^{-1}$. They used that sample to
determine the H$\alpha$ and SFR density luminosity functions for the
local Universe. We proceeded here in a similar manner to estimate the
[OII]$\lambda$3727 luminosity function.  Instead of adding artificial
galaxies, we chose a final limiting flux slightly fainter than the one
provided by the V/Vmax method and then we included all galaxies having
non-zero flux in {\sc [OII]}$\lambda$3727.  With the goal of
quantifying the goodness of V/Vmax results, we compared with the
maximum-likelihood parametric fit \citep[STY,][]{1991ApJ...372..380Y}
and the nonparametric step-wise maximum-likelihood
\citep[SWML,][]{1988MNRAS.232..431E} methods.

The resulting Schechter best fitting parameters as provided
by the STY method for a limiting {\sc [OII]} flux of 
$2.2\times10^{-14}$ erg cm$^{-2}$ s$^{-1}$ are:
$\alpha-1.21\pm0.21$, $\phi^*=10^{-2.98\pm0.19}$ {\rm Mpc}$^{-3}$, and 
$L^*=10^{40.92\pm0.13}$ {\rm erg} {\rm s}$^{-1}$.

The number densities for each luminosity interval obtained from the
V/Vmax (both observed and extinction corrected) are tabulated in
Table~\ref{tab-phi}.  Both fit and values, within errors, are almost
identical to the ones obtained by both the V/Vmax and SWML methods.
The densities were corrected for the 8\% of UCM galaxies for which
there is no [OII] data available.  The luminosity function errors are
those obtained considering a Gaussian distribution of $\sigma$ equal
to the square root of the number of galaxies in each $\log L({\rm
OII})$ bin.

\placetable{tab-phi}

Any comparison with previous [OII] luminosity functions has to be done
with caution, given that different measurements come from samples
built with very different selection criteria. In Figure \ref{fig-lf}
we have plotted our {\sc [OII]} observed luminosity function for the
local Universe, in comparison with the one by
\citet{1998ApJ...504..622H} for the Caltech Faint Galaxy Redshift
Survey (CFGRS), a sample of galaxies at intermediate redshift
($0.35<z<1.5$).  The [OII] luminosity density of the Universe was
obviously higher in the past.

Given that we have E(B--V) for every object but seven, we
corrected each object luminosity as:
\begin{eqnarray}
 L_{corr}^{OII} \; = \; EW_{obs}^{OII} \; L_{obs}^{c,OII}  \; 
10^{0.4 \: (A_{3727} \: E_l(B-V))} 
\end{eqnarray}
where $A_{3727} = \frac{f(3727)}{0.295} = 4.81$ \citep{1989agna.book.....O}.  

The resulting Schechter best fitting parameters as provided by the STY
method for a limiting flux
of $4.0\times10^{-14}$ erg cm$^{-2}$ s$^{-1}$ are:
$\alpha=-1.17\pm0.08$, $\phi^*=10^{-3.71\pm0.16}$ {\rm Mpc}$^{-3}$,
and $L^*=10^{42.66 \pm 0.17}$ {\rm erg} {\rm s}$^{-1}$. Densities are in Table~\ref{tab-phi} (again V/Vmax and SWML results are similar).

\citet{2000MNRAS.312..442S} published the luminosity function for a
sample of 273 galaxies in the range $z=0.01-0.3$ selected by their
emission in the ultraviolet.  Their {\sc [OII]}$\lambda$3727
luminosities were obtained from follow-up spectroscopy and were
extinction-corrected. A Schechter fit provides the following
parameters: $\alpha=-1.59\pm0.12$, $\phi^*=10^{-2.82\pm 0.18}$
Mpc$^{-3}$, and $L^*=10^{41.96\pm0.09}$ erg s$^{-1}$.  The differences
have to be understood as coming different galaxy populations, with
more abundant but less luminous galaxies in the UV-selected sample.

\section{The evolution of the {\sc [OII]}$\lambda$3727 luminosity density}

One of the major uncertainties when analyzing SFR densities at
different redshifts arises when results obtained with several 
tracers have to be compared. This is why it is so important to obtain the
evolution of the luminosity densities for each SFR tracer.
Our local {\sc [OII]}$\lambda$3727 luminosity density, when combined
with similar results from deep samples up to $z\sim1.5$, allow us to
sketch the evolution of the SFR density as traced by the observed {\sc
[OII]}luminosity.  Given that the luminosity
function is well fitted as a Schechter function, and $\alpha \leq
-2$, $\phi(L)$ can be integrated for the whole range of luminosities:
\begin{eqnarray}
L_{tot} \; = \; \int^{\infty}_0 L \: \phi(L) \: dL \; = \; 
\phi^* \: L^* \: \Gamma(2+\alpha)
\end{eqnarray}
For the observed luminosities, the total {\sc [OII]}
luminosity density is
$10^{38.01\pm0.15}\,$erg$\,$s$^{-1}\,$Mpc$^{-3}$ in the Local Universe
($z\lesssim0.045$) for star-forming galaxies with $EW({\rm
H}\alpha+[{\rm NII}])\gtrsim20$\AA.  When the effect of the extinction is
considered, the total {\sc [OII]} extinction-corrected luminosity
per unit volume corresponds to
$10^{39.00\pm0.11}\,$erg$\,$s$^{-1}\,$Mpc$^{-3}$. The extinction
correction amounts for a total of one dex. This luminosity density
is similar to the one we could estimate from the B band luminosity
function published by citet{1997ApJ...475..502G} for the UCM sample. It
is also comparable to the luminosity density we would obtain if we apply
an average [OII]$\lambda$3727/H$\alpha$ factor to the extinction-corrected 
H$\alpha$ luminosity density measured by \citet{1995ApJ...455L...1G}.

In Figure \ref{fig-madauOII} we have plotted all the {\sc [OII]} luminosity
densities published in the literature. The CFGRS sample 
 already revealed a strong evolution from z=1 to the low
redshift universe. This evolution is in agreement with the 
Canada-France Redshift Survey results \citep{1997ApJ...481...49H}. 
Unfortunately, these surveys only provide with observed
luminosities. We have also plotted both the observed and the
extinction-corrected value for the \citet{2000MNRAS.312..442S} sample.

The right axis of Figure~\ref{fig-madauOII} shows how the $L({\rm
[OII]})$ scale transforms into a SFR density scale using the nominal
transformation provided by \citet{1998ARA&A..36..189K}:
$\frac{SFR}{{\rm M}_\odot{\rm yr}^{-1}}=1.4\times10^{-41} L_{\rm
[OII]}$ {\rm (erg s$^{-1}$).  Our total extinction-corrected {\sc
[OII]} luminosity density translates into a SFR density of
$0.014\pm0.003\,{\rm M}_{\odot}\,$yr$^{-1}\,$Mpc$^{-3}$.

Assuming that the SFRD evolves with redshift as $(1+z)^n$, the
combination of the UCM local value with the CFGRS data implies a
$n\approx4$.  Such strong evolution is similar to the one obtained
using H$\alpha$ as SFR tracer for the same redshift range.  However,
we want to stress that the selection effects and calibrations for each
sample were different and some caution is necessary when interpreting
these quantitative results.

\acknowledgments

Valuable discussions with A. Alonso-Herrero, P.G. Pérez-González and
S. Pascual are gratefully acknowledged.  We would also like to thank
the anonymous referee for his/her useful suggestions that improved this
paper.  This work was supported in part by the Spanish Plan Nacional
de Astronomía y Astrofísica under grant AYA2000-1790.

\bibliographystyle{apj}
\bibliography{referencias}

\begin{deluxetable}{ccccc}
\tablewidth{0pc}
\tablecaption{{\sc [OII]}3727 Luminosity Function for the UCM Survey}
\tablehead{
\colhead{} &
\multicolumn{2}{c}{Observed LF} & \multicolumn{2}{c}{Corrected LF} \\
\colhead{log L([OII]3727)} & \colhead{log($\Phi$)} & \colhead{Number of} 
& \colhead{log($\Phi$)} & \colhead{Number of} \\
\colhead{[erg s$^{-1}$]} & \colhead{[\# Mpc$^{-3}$ per} & \colhead{galaxies}
& \colhead{[\# Mpc$^{-3}$ per} & \colhead{galaxies} \\
\colhead{} & \colhead{logL interval]} & \colhead{}
& \colhead{logL interval]} & \colhead{}
}
\startdata
 $39.0$ & \nodata         & \makebox[0.5cm][r]{$0$} &\nodata        &\makebox[0.5cm][r]{$0$}\\   
 $39.4$ & $ -2.63\pm0.22$ & \makebox[0.5cm][r]{$4$} &$-2.48\pm0.43$ &\makebox[0.5cm][r]{$1$}\\  
 $39.8$ & $ -2.82\pm0.14$ & \makebox[0.5cm][r]{$10$}&\nodata        &\makebox[0.5cm][r]{$0$}\\  
 $40.2$ & $ -2.89\pm0.08$ & \makebox[0.5cm][r]{$33$}&$-3.06\pm0.14$ &\makebox[0.5cm][r]{$9$}\\   
 $40.6$ & $ -3.08\pm0.06$ & \makebox[0.5cm][r]{$60$}&$-3.30\pm0.10$ &\makebox[0.5cm][r]{$18$}\\ 
 $41.0$ & $ -3.61\pm0.09$ & \makebox[0.5cm][r]{$22$}&$-3.58\pm0.09$ &\makebox[0.5cm][r]{$23$}\\ 
 $41.4$ & $ -4.25\pm0.19$ & \makebox[0.5cm][r]{$5$} &$-3.47\pm0.08$ &\makebox[0.5cm][r]{$33$}\\ 
 $41.8$ & \nodata         & \makebox[0.5cm][r]{$0$} &$-3.59\pm0.09$ &\makebox[0.5cm][r]{$25$}\\ 
 $42.2$ & \nodata         & \makebox[0.5cm][r]{$0$} &$-3.88\pm0.12$ &\makebox[0.5cm][r]{$13$}\\ 
 $42.6$ & \nodata         & \makebox[0.5cm][r]{$0$} &$-4.69\pm0.31$ &\makebox[0.5cm][r]{$2$}\\ 
 $43.0$ & \nodata         & \makebox[0.5cm][r]{$0$} &$-4.51\pm0.25$ &\makebox[0.5cm][r]{$3$}\\ 
\enddata					       
\label{tab-phi}					       
\end{deluxetable}				       
						       
\clearpage					       
						       
\onecolumn					       
						       
\begin{figure}					       
\epsscale{0.75}					       
\plotone{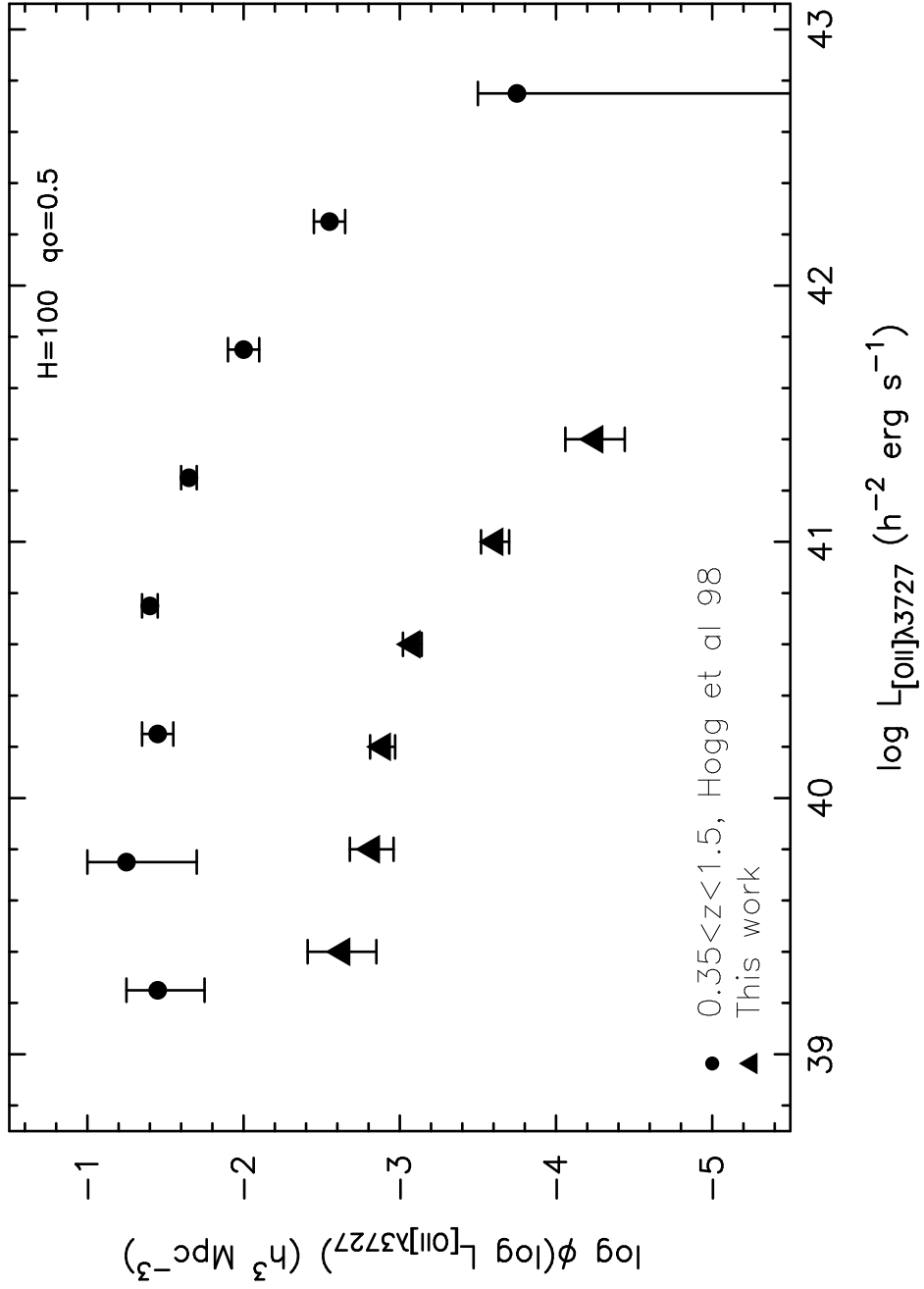}				       
\caption{Observed {\sc [OII]}$\lambda$3727 luminosity functions for both 
the UCM (filled triangles) and the CFGRS (filled circles) samples.}
\label{fig-lf}
\end{figure}

\clearpage

\begin{figure}
\epsscale{0.6}
\plotone{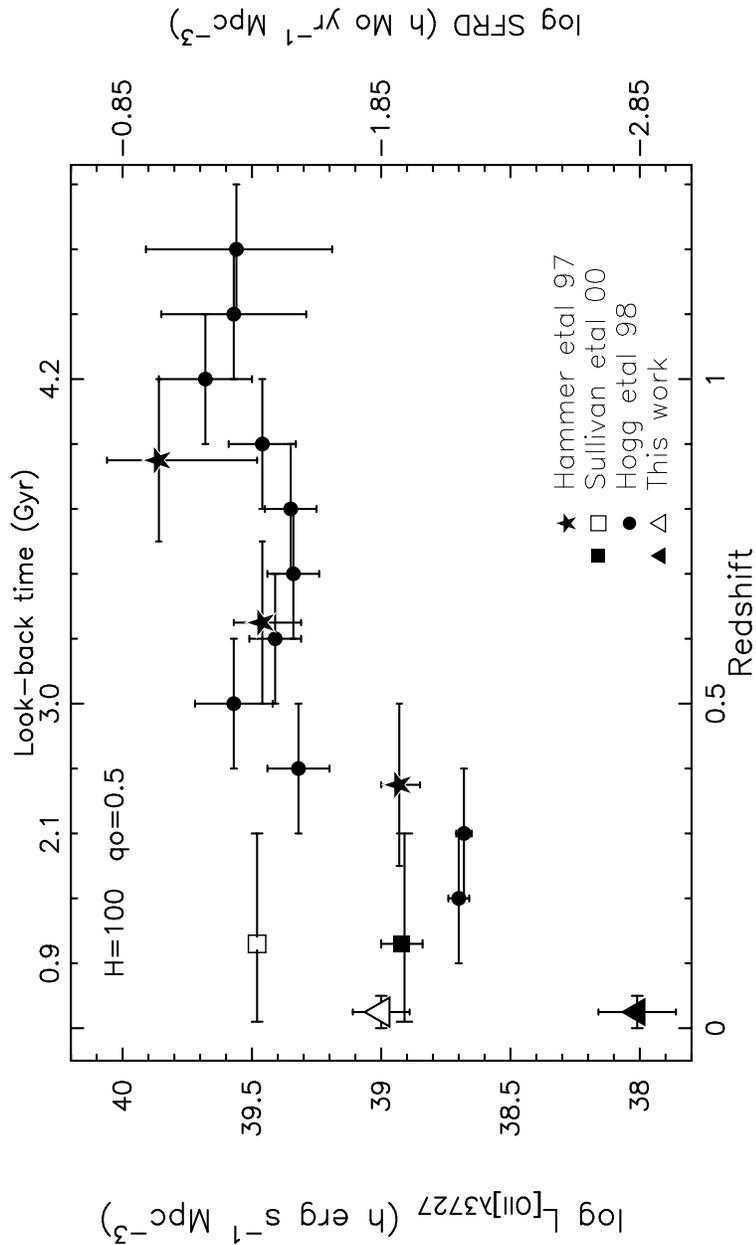}
\caption{The evolution of the {\sc [OII]}$\lambda$3727 luminosity
density of the Universe.  Both the observed and extinction-corrected
values for the UCM sample are represented as filled and open triangles.  Filled
circles correspond to \citet{1998ApJ...504..622H}, filled stars are for the
\citep{1997ApJ...481...49H} values, 
and the filled and open squares are for the observed and extinction corrected 
\citet{2000MNRAS.312..442S} values.}
\label{fig-madauOII}
\end{figure}

\end{document}